\documentstyle[12pt,doublespace]{article}
\begin{document}
\pagestyle{empty}
\vspace{0.20in}
\begin{center}
{\Large Charge Transfer Excitons and Possible Exitonic Pairing in
the Extended Three Band Hubbard Model.}
\end{center}
\vspace{0.40in}
\begin{center}
C. Vermeulen$^{1}$, W. Barford$^{1}$ and  E. R. Gagliano$^{2}$
\end{center}
\begin{center}
1. Department of Physics, The University of
Sheffield, Sheffield, S3 7HR, United Kingdom.\\
2. Centro Atomico Bariloche, 8400 Bariloche, Argentina. \\
\end{center}
\vspace{0.40in}
\begin{center}
Abstract
\end{center}
Exact diagonalisations of the extended Hubbard model are performed.
In the insulating regime it is shown that the nearest neighbour copper-oxygen
repulsion, $V$,
leads to Frenkel
 excitons in the charge transfer gap at values of
$V$ of the order of copper-oxygen hybridisation, $t$.
 In the metallic regime it is shown that the static charge-transfer and
 density-density correlation functions diverge as a function of $V$,
 indicating a charge-tansfer instability and phase separation.  This is
 accompanied by a softening of the $q \rightarrow 0$ mode of the dynamic
 correlation functions which is associated
 with the excitonic excitations responsible for the superconducting
correlations
 observed in the proximity of the phase separation boundary of ref.
\cite{cv95b}.

\vspace{0.5in}

\vspace{0.5in}
\noindent
March, 1995.

\vfill\eject

The charge-transfer fluctuations $Cu^{2+} \rightarrow Cu^+$ and $O^{2-}
\rightarrow O^-$ coupled to the nearest neighbour copper-oxygen
repulsion, $V$, leads to a variety of interesting phenomena.
For example, in the insulating-stochiometric limit it leads to the formation
of charge-transfer excitons \cite{yu91}, \cite{cv95a}, for which there is some
 experimental
evidence from Raman \cite{lui93} and optical \cite{perk93} experiments.
In the metallic regime the softening of the charge-transfer excitations
leads to a charge-transfer instability and phase separation, which suggests
the possibility of an excitonic pairing mechanism \cite{varma87},
\cite{Nun88}.  This instability is
also related to the pinning of the Fermi energy on hole doping, and to the
creation of new states in the charge-transfer gap \cite{wb95}.  Such behaviour
 is
qualitatively similar to the observations from photoemission \cite{jwa}
and $O$ $ 1s$ spectroscopy \cite{rom}.

The three band extended Hubbard model, which best describes this behaviour,
has been studed by a number of authors.  Littlewood {\it et al.}
\cite{littlewood89},
following the initial suggestion of Varma {\it et al.} \cite{varma87},
studied the
charge-transfer excitations in the weak coupling limit, via the RPA.  This
work was followed up by Bang {\it et al.} \cite{bang93} in the weak coupling
 limit,
 and by Raimondi {\it et al.} \cite{Rai93} in the strong coupling limit.
These analyses all
 indicated superconducting correlations, mediated by excitonic excitations,
 close to the charge-transfer instability.
  Barford and Long \cite{wb93} performed a unitary transformation to second
 order in
 the copper-oxygen hybridisation and found superconductivity and phase
 separation.  Finally, Sano and Ono \cite{san93}, Vermeulen {\it et al.}
 \cite{cv95b} and Stechel {\it et al.} \cite{stec95} used
 the mapping onto a Luttinger liquid to show that the one dimensional model
 exhibits superconducting correlations and anomalous flux quantisation in the
 proximity of the phase separation boundary.

It is the purpose of this letter to study in detail the dynamics of the
 charge-transfer excitations as a function of doping and $V$ by solving exactly
 finite size clusters in one dimension.  We will study the charge-transfer
 excitons in the insulating regime. By performing finite size calculations we
 will
compare the exact results to the predictions of the variational theory
of Vermeulen and
Barford \cite{cv95a}.  Next, by studying the dynamic charge-transfer and
density-density correlation functions in the metallic regime we will relate the
softening of the $q \rightarrow 0$ mode and the divergence of the static
correlation functions to the superconducting correlations and phase separation
instabilities of the copper-oxide chain  phase diagram of ref. \cite{cv95b}.

Our model for the copper-oxide chain consists of two atoms per unit cell.
Neglecting the oxygen-oxygen hybridisation and considering the Coulomb
interaction up to first-nearest neighbours, the copper-oxide chain is described
 by the two band model Hamiltonian
\begin{eqnarray}
H &=& -t \sum_{<ij>\sigma} (d^{\dagger}_{i\sigma}
p_{j\sigma} + h.c.)
 + {\Delta \over 2} \sum_{ij\sigma}(p^{\dagger}_{j\sigma}p_{j\sigma}
    -d^{\dagger}_{i\sigma}d_{i\sigma}) \nonumber \\
         &+&  U_{d} \sum_{i} d^{\dagger}_{i\uparrow}d_{i\uparrow}
        d^{\dagger}_{i\downarrow}d_{i\downarrow}
        +  U_{p} \sum_{j} p^{\dagger}_{j\uparrow}p_{j\uparrow}
        p^{\dagger}_{j\downarrow}p_{j\downarrow}
  +  V \sum_{<ij>\sigma\sigma^\prime} d^{\dagger}_{i\sigma}d_{i\sigma}
             p^{\dagger}_{j\sigma^\prime}p_{j\sigma^\prime},
\end{eqnarray}
\noindent
where $i$ and $j$ are copper and oxygen sites respectively,
$<ij>$ represents nearest neighbours and the operator
$d^{\dagger}_{i\sigma}~~(p^{\dagger}_{j\sigma})$ creates a $Cu$ $(O)$ {\it
hole}
with spin $\sigma$.  $\Delta$ is the charge-transfer energy, $U_d$ $(U_p)$ is
the copper (oxygen) Coulomb repulsion and $V$ and $t$ are the copper-oxide
Coulomb repulsion and hybridisation, respectively. We will denote $N_{p}$ as
 the number of particles and  $N_s$ as the number of sites. Our choice of model
 parameters will be
those relevant to the high  temperature superconductors. Thus we take
 $U_d=8t$, $\Delta=2-4t$ and $U_{p}=0-4t$ where $t\sim1.5eV$ \cite{Tun91}.
 However, as we
 are interested in the r$\hat{\mbox{o}}$le played by the nearest neighbour
 interaction, $V$, this will be left as a free parameter.

Firstly we examine the model in the insulating phase at a hole density of
$n=0.5$.  At this density the system is a charge transfer insulator for all
$\Delta\neq0$, and the ground state is a spin density wave. The majority of
the charge resides on the copper sites, forming a N\'eel state.  The exciton
is identified as the lowest frequency peak, $E_{exc}$, of the dynamical
current-current correlation function, $J(w)$, when it appears inside the
charge transfer gap, $E_{gap}$.  $J(w)$ is defined by
\begin{eqnarray}
J(w)=-\frac{1}{\pi}Im(G^{R}(w)), \nonumber
\end{eqnarray}
where
\begin{eqnarray}
\hspace{1cm} G^{R}(w) &=& F.T.<\psi_{0}\mid
j^{\dagger}(t)j(0)\mid\psi_{0}> \nonumber\\
&=& <\psi_{0}\mid j^{\dagger}\left(\frac{1}{H-(E_{0}+w)+i\eta}\right)j
\mid\psi_{0}>_{\eta\rightarrow0}.
\end{eqnarray}
$G^{R}(w)$ is the retarded Green function and $j$ is the
 current operator defined as $j=-i\sum_{l\sigma}$ $(c^{\dagger}_{l\sigma}
c_{l+1\sigma}-c^{\dagger}_{l+1\sigma}c_{l\sigma})$.  $\{\mid\psi_{0}>, E_0\}$
 are the ground state eigensolutions. In practice we choose a finite value of
 $\eta$ to broaden the peaks.  The Green function is calculated via the
 continued fraction technique \cite{Gag89}. The energy of the gap is defined
 by $E_{gap}= E(N_p+1) + E(N_p-1) -2E(N_p)$. Figure 1 shows a typical profile
 for the excitonic spectrum using a $12$ site chain, and the values of
 $U_d=8t$, $U_p=0$, $\Delta=2t$ and $V=2t$. The values of $V_{crit}$, the
 point at which $E_{gap}=E_{exc}$, are shown in tables (1) and (2) for 8, 12
 and 16 site chains for $\Delta=2t$ and $4t$, respectively.  The values for
 $V_{crit}$ at $N=\infty$ are found by finite size scaling, whereby
 $E^{N_{s}}_{gap}(V)$ and $E^{N_s}_{exc}(V)$ are plotted against
 $\frac{1}{N_s^2}$ and the straights lines are extrapolated to
 $\frac{1}{N_s^2}=0$.  Notice that $V_{crit}$, which is of the order $t$, is a
 decreasing function of $\Delta$ and $U_p$.

The exciton may be pictured as a hole that has been excited from the lower
Hubbard band (or conduction band), which is predominately copper in character,
to the valence band, which is predominately oxygen in character. The real
space picture is therefore of a hole on a copper site hopping onto a
neighbouring oxygen site. This effectively leaves an `electron' on the copper
site and a hole on the oxygen site, and due to the Coulomb repulsion between
neighbouring sites this leads to an effective attraction between the electron
and the hole. This attraction reduces the energy to a point below the charge
transfer gap and causes the formation of a tightly bound `electron-hole' pair,
or a Frenkel exciton. The energy in forming the exciton is a balance between
the Coulomb attraction of the `electron-hole' pair and the kinetic energy
loss.  The former is driven by $V$, whereas the latter is determined by $t$.
Hence, when $V$ is of the order of $t$ we expect excitons to exist within the
charge transfer gap.

As a more quantitative study, we have extended
 the variational estimate by Vermeulen and Barford \cite{cv95a} of $V_{crit}$
 for the two dimensional copper-oxide plane.  This was performed  via a
 canonical transformation up to O($t^2$) in the hopping $t$ of eqn(1),
 assuming strong coupling ($U_d \rightarrow \infty$). Figure 2 shows
 $\tilde{V}_{crit}$ against $\tilde{\Delta}$ for the one dimensional infinite
 chain and several values of $U_p$  (where $\tilde{V}=\frac{V}{\tilde{t}}$,
 $\tilde{\Delta}= \frac{\Delta}{\tilde{t}}$ and
 $\tilde{t}=\frac{t^2}{\Delta}$).  Notice that for large $\tilde{\Delta}$
 (where the calculation becomes asymtopically rigorous) $\tilde{V}_{crit}$
 {\it decreases} as a function of $U_p$, and is a constant function of
 $\tilde{\Delta}$, {\it i.e.} $V_{crit}$ decreases with $\Delta$.  This is in
 agreement with the numerical results of tables (1) and (2), and  arises from
 the fact that increasing the oxygen repulsion increases the single-particle
 gap, but has little effect on the exciton energy.  In contrast, the
 variational  estimate of ref\cite{cv95a} indicates that in two dimensions
 $U_p$ has no effect on $V_{crit}$ for large $\tilde{\Delta}$.  We also
 confirm the validity of the variational estimate by  comparing $V_{crit}$
 with the finite size numerical results in table (2), which indicates good
 agreement.

We now turn to the r$\hat{\mbox{o}}$le that the charge transfer excitations
 may play in the doped phase of the cuprate superconductors. From previous
work \cite{cv95b} this model is well known to undergo phase separation within
a hole density range of $0.5<n<1.0$ and large enough $V$. There is also strong
evidence that the system becomes superconducting in the proximity of the phase
separation boundary.  We intend to investigate the r$\hat{\mbox{o}}$le played
by correlated particle-hole excitations near the Fermi surface in these two
physical effects.

It is difficult to identify excitonic peaks from the current-current
 correlation function in the metallic regime.  Instead, it is more convenient
 to study the dynamic density-density and  charge-transfer correlation
functions ({\it i.e.} susceptibilities) at low momenta. The dynamic
 charge-transfer correlation function, in particular, focuses on local
 intra-cell charge fluctuations of copper and its surrounding oxygens. Such a
local fluctuation in charge can be considered excitonic in character, and at
low energy and momenta these will correspond to excitations near the Fermi
 surface. The dynamic charge-transfer correlation function is defined as
\begin{equation}
 \chi^{\delta}_{q}(w)=F.T.<\delta^{\dagger}_{q}(t)\delta_{q}(0)>,
\end{equation}
where
\begin{equation}
\delta_q=\frac{1}{\sqrt{N_s}}\sum_{l}(-1)^l n_l e^{iqx_l}.
\end{equation}
The dynamic density-density correlation function, $\chi^n_q(w)$, is
defined by replacing $\delta_q$ by $n_{q}=\sum_{k}c_{k+q}^{\dagger}c_{k}$. In
terms of our lattice Hamiltonian $n_{q}$ is defined as
 $\frac{1}{\sqrt{N_s}}\sum_{l}(n_l-<n>)e^{-iqx_l}$. This function
 examines charge fluctuations across the whole lattice. From a
 thermodynamical argument it can be shown that the compressibility, $\kappa$,
 is related to the static density-density correlation function in the long
wavelength
 limit,
 $C^n_{q\rightarrow0}$,  by
\begin{equation}
\kappa=\frac{\beta}{V\rho^3}C^n_{q\rightarrow0}.
\end{equation}
$C^n_{q\rightarrow0}$ is the integrated dynamic correlation function
 $(= \int \chi^n_{q\rightarrow0}(\omega) d\omega)$, $\rho$ is the
density, $\beta$ is the inverse temperature and $V$ is the volume. So a
 divergence of the static density-density correlation function in the limit
 $q\rightarrow0$ is directly related to a divergence of $\kappa$, and hence to
 phase separation.  Moreover, the divergence of the static correlation function
  should be accompanied by a soft mode.  Figures 3(a) and (b) show the
position of the lowest peak in the dynamic charge-transfer correlation
 function and the integrated weight for $q=\frac{2_\pi}{N_s}$ ({\it i.e.} the
 finite size $q \rightarrow 0$ limit) on a 12 site chain at a filling of
 $8/12$ and $10/12$, respectively.  As the nearest neighbour repulsion is
 increased the static charge-transfer correlation function rapidly increases
 and diverges at about $V \sim  2.0t$ for 8 holes and $V \sim 1.5t $ for 10
 holes.  (The divergence is taken to be at the point of inflection, as a
 finite size system cannot have a truly diverging susceptibility.)  This is
 accompanied by the softening of the diverging $q \rightarrow 0$, $\omega
 \rightarrow 0$ mode.  (All other $q$ modes stiffen as $V$ is increased.)
 Likewise, the static density-density correlation function also diverges.  The
 frequency of the density-density soft mode is identical to the
 charge-transfer soft mode, as charge-transfer and density fluctuations are
 coupled. This arises from the strong Coulomb interactions on copper which
 means that local charge fluctuations will tend not to conserve local charge
 and so these excitations are coupled to non-local charge fluctuations. The
 divergence of the static correlation functions is consistent with the
 divergence of the discrete compressibility at $V \simeq 2.0t$ and $1.6t$ for
 8 and 10 holes, respectively, as shown in the phase diagram of ref.
 \cite{cv95b}.  Moreover, it is reasonable to associate the softening of the
 $q \rightarrow 0$ charge-transfer excitations with the indications of
 superconducting correlations from both the Luttinger liquid charge exponent,
 $K_{\rho}$, exceeding unity and to the onset of anomalous flux quantisation
 \cite{cv95b}.

The effect of $U_p$ is to suppress strongly both the static density-density
and charge-transfer correlation functions. At $n=\frac{8}{12}$ and $V=3t$
$C^n_{q \rightarrow 0}$ drops from $0.22$ at $U_p=0$ to $0.09$ at $U_p=2t$.

In conclusion, we have shown that the nearest neighbour repulsion leads
naturally to Frenkel excitons in the charge-transfer gap at values of
$V_{crit}$ of the order of $t$ in the insulating regime. Further, we have found
 that in one dimension increasing $U_p$ has the effect of decreasing the value
 of $V_{crit}$, and we have given a qualitative argument to explain this
effect.
 These numerical results are in good agreement with the analytical results of
 \cite{cv95a}.

In the metallic regime we have shown that the static charge-transfer and
 density-density correlation functions diverge as a function of $V$,
 indicating a charge-tansfer instability and phase separation.  This is
 accompanied by a softening of the $q \rightarrow 0$ mode in the dynamic
 correlation
 functions which we associate
 with the excitonic excitations responsible for the superconducting
correlations
 observed in the proximity of the phase separation boundary of ref.
\cite{cv95b}.

\vspace{1 cm}

\begin{center}
Acknowledgements\\
\end{center}
We thank the SERC (United Kingdom) for the provision of a Visiting Research
Fellowship (ref. GR/H33091). W.B. also acknowledges a grant from the
University of Sheffield research fund. C.J.V. is supported by a University of
Sheffield scholarship.  We thank  R. Bursill and M. Grilli for stimulating
discussions.

\vfill\eject

\pagebreak
     Table 1.
     \begin{tabular}{|c||c|c|c|c|c|}
     \hline
     $\Delta=2t$ & \multicolumn{5}{|c|}{Number of lattice site, $N_s$} \\
      \cline{2-6}
     $U_d=8t$ & 8 & 12 & 16 &  & $\infty$ \\
     \hline
     $U_p=0$ & $2.06t$ & $1.7t$ & $1.54t$ & \ldots & $1.29t$ \\
     \hline
     $U_p=4t$ & $1.57t$  & $1.22t$ & $1.09t$ & \ldots & $0.88t$ \\
     \hline
     \end{tabular}
     \\
     \vspace{1in}
     \\
     Table 2.
     \begin{tabular}{|c||c|c|c|c|c|c|}
     \hline
     $\Delta=4t$ & \multicolumn{6}{|c|}{Number of lattice site, $N_s$} \\
      \cline{2-7}
     $U_d=8t$ & 8 & 12 & 16 &  & $\infty$ & Analytic \\
     \hline
     $U_p=0$ & $2.02t$ & $1.54t$ & $1.38t$ & \ldots & $0.94t$ & $1.22t$ \\
     \hline
     $U_p=4t$ & $1.82t$  & $1.31t$ & $1.16t$ & \ldots & $0.77t$ & $1.05t$ \\
     \hline
     \end{tabular}

\begin{figure}
\begin{center}
{\bf Figure Captions}
\end{center}

\vspace{0.3in}

{\bf Figure 1.} The current-current response function at $U_d=8t$, $U_p=0t$,
 $\Delta=2t$ and $V=2t$ for the stoichiometric 12 site chain. $E_g$ is the
 charge
 transfer gap energy and $E_e$ identifies the exciton energy.

\vspace{0.3in}

{\bf Figure 2.} $\tilde{V}_{crit} (=V_{crit}/\tilde{t})$ versus $\tilde{\Delta}
(=\Delta/\tilde{t})$, where $\tilde{t}=t^2/\Delta$, for the one dimensional
 chain, using the theory of ref. \cite{cv95a}, in the $U_d=\infty$ limit.  The
 exciton is a bound state above the lines.

\vspace{0.3in}

{\bf Figure 3.} (a) The  $q=\frac{\pi}{6}$ static charge-transfer (diamonds)
and
 density-density (solid squares) correlation functions, and the position of the
 lowest frequency mode of the dynamic correlation functions (open squares) as a
 function of $V/t$ for 8 holes in a 12 site chain.
The frequency is in units of $t$.  Notice the different normalisations for the
correlation functions. $U_d=9t$, $U_p=0$ and $\Delta=2t$.
(b)  The same as (a) with 10 holes.

\vspace{0.3in}

{\bf Table 1.} Values of $V_{crit}$ for various system sizes.

\vspace{0.3in}

{\bf Table 2.} Values of $V_{crit}$ for various system sizes.
The analytic result is in the $U_d=\infty$ limit.

\end{figure}

\end{document}